\documentstyle[12pt,epsf]{article}
\renewcommand{\bar}[1]{\overline{#1}}

\begin{document}

\begin{flushright}
USM-TH-164\\
CPT-2004/P.134\\
\end{flushright}
\bigskip\bigskip

\centerline{\large \bf Transverse single spin asymmetries in photon production}

\vspace{22pt}

\centerline{\bf {
Ivan Schmidt\footnote{Electronic address: ivan.schmidt@usm.cl}$^{a}$,
Jacques Soffer\footnote{Electronic address: soffer@cpt.univ-mrs.fr}$^{b}$,
Jian-Jun Yang\footnote{deceased}$^{a,c}$}}

{\centerline {$^{a}$Departamento de F\'\i sica, Universidad
T\'ecnica Federico Santa Mar\'\i a,}}

{\centerline {Casilla 110-V,
Valpara\'\i so, Chile}}

{\centerline {$^{b}$ Centre de Physique Th\'eorique, UMR 6207
\footnote{UMR 6207 is Unit\'e Mixte de Recherche
du CNRS et des Universit\'es Aix-Marseille I,\\
 Aix-Marseille II et de
l'Universit\'e du Sud Toulon-Var - Laboratoire affili\'e à la FRUMAM.},}}
{\centerline {CNRS-Luminy, Case 907
F-13288 Marseille Cedex 9 - France}}

{\centerline {$^{c}$Department of Physics, Nanjing Normal
University,}}

{\centerline {Nanjing 210097, China}}

\vspace{10pt}
\begin{center}
{\large \bf Abstract}
\end{center}

Transverse single-spin asymmetries (SSA) in inclusive reactions
are now considered to be directly related to the transverse
momentum ${\bf k}_{T}$ of the fundamental partons involved in the
process. We find  that the ideal probe to extract information on
the gluon Sivers function is the transverse SSA of prompt photon
production $p p^{\uparrow} \to \gamma X$, at large $p_T$. The
following related processes, $p p^{\uparrow} \to \gamma + jet +
X$, $p p^{\uparrow} \to \gamma^* + X \to \mu^+ \mu^- + X$ and
$\bar{p} p^{\uparrow} \to \gamma + X$ are also briefly discussed.

\newpage

At present there is a wealth of experimental observations of
single spin asymmetries (SSA) in many different processes. Large
SSA have been measured in $p p^{\uparrow} \rightarrow \pi X$,
where one proton is transversely polarized, and in which the
produced pion prefers to come out, either to the right or to the
left of the plane formed by the beam direction and the proton
polarization vector, depending on its charge. This effect was
first observed at FNAL more then ten years ago, in experiments
done by the E704 Collaboration \cite{E704}, at center-of-mass
(c.m.) energy $\sqrt{s} \sim 20~\mbox {GeV}$. It occurs also at
$\sqrt{s} = 200 ~\mbox {GeV}$, as observed recently for $\pi^{0}$
production by the STAR Collaboration \cite{STAR}, in the first
spin run at BNL-RHIC. Although the data appear to have very little
energy dependence, a careful study of the unpolarized cross
section leads to conclude that the SSA, in these two energy
regimes, may have two different dynamical origins \cite{BS}. Several
SSA have been also measured in hyperon (and antihyperon) inclusive
production $pN\rightarrow Y^{\uparrow}X$, at various
energies~\cite{LP}, but a suitable detailed interpretation of
these rich polarization data is still missing. Moreover, recently
an azimuthal asymmetry has been also observed in semi-inclusive
deep-inelastic scattering (SIDIS) $ {\it l}p^{\uparrow}
\rightarrow {\it l}\pi X$, for targets polarized transversely
($A_{UT}$) and longitudinally ($A_{UL}$) relative to the direction
of the unpolarized incoming lepton beam direction
~\cite{SMC,hermes}.

Although these SSA are not yet fully understood, they are expected
to give valuable information on the orbital angular momenta of
quarks and gluons inside the hadron. Furthermore, they provide us
with an understanding of QCD at the amplitude level, which comes
from the fact that the SSA is proportional to the interference of
a spin flip and a non spin flip amplitude, out of phases.
Therefore in perturbation theory such an interference effect,
which requires an imaginary part, is generated at the one loop
level. The interference is between wave functions with angular
momenta $J_z = \pm 1/2$ and hence contains information on the
parton´s orbital angular momenta~\cite{BHS1}. Moreover, the
required matrix element measures the spin-orbit correlation
$\vec{S} \cdot \vec{L}$ within the target hadron's wavefunction,
the same matrix element which produces the anomalous magnetic
moment of the proton, the Pauli form factor, and the generalized
parton distribution $E$ which is measured in deeply virtual
Compton scattering.

In practice, essentially two mechanisms have been proposed in
order to explain the SSA. The first one is to generalize the
parton distribution functions by considering distributions that
depend on the transverse momenta ${\bf k}_{T}$ of these partons,
and the second is to take into account higher twist operators
\cite{QS92}. Recently it was shown that there is a direct relation
between these two approaches, so in fact they are expected to
produce very similar effects. In the case of the ${\bf k}_{T}$
dependent distribution functions, the SSA can be produced either
by quark distributions, which is called the Sivers effect
\cite{DS}, proposed long time ago, or by quark fragmentation
functions, which is called the Collins effect \cite{JC}. For some
time it was thought that the Sivers function vanished, but this
was shown not to be the case in an explicit simple model
calculation~\cite{BHS1}.

In general both the Sivers and the Collins effects will be present
in a specific reaction, although there are some cases in which
only one of them contributes. For example, the Collins effect is
the only mechanism that can lead to asymmetries $A_{UT}$ and
$A_{UL}$, defined above. On the other hand, it does not appear in
some electroweak interaction processes, where there is only the
Sivers effect. In this paper we will concentrate on the Sivers
function, whose existence was proved by considering final state
interactions in a diquark model \cite{BHS1,BSY03}. The diquark
model can only predict the Sivers function for the valence quarks,
and it is also of interest to calculate it for sea quarks or for
gluons. In fact, the gluon Sivers function was mentioned for the
first time in Ref.~\cite{SS03}, and only recently it was also
considered in jet correlations \cite{BV} and in $D$ meson
production \cite{ABALM} in $p^{\uparrow}p$ collisions. Just as the
quark Sivers function is related to the hadron´s anomalous
magnetic moment, the gluon Sivers function is connected with the
gluon´s contribution to the same anomalous magnetic moment, a
quantity which in general is difficult to obtain.

The direct photon production in $pp$ collisions can provide a
clear test of short-distance dynamics as predicted by perturbative
QCD, because the photon originates in the hard scattering
subprocess and does not fragment, which immediately means that the
Collins effect is {\it not} present. This process is very
sensitive to the gluon structure function, since it is dominated
by the quark-gluon Compton subprocess in a large photon transverse
momentum range. Prompt-photon production, $pp (p\bar{p}) \to
\gamma X$, has been a useful tool for the determination of the
unpolarized gluon density and it is considered one of the most
reliable reactions for extracting information on the polarization
of the gluon in the nucleon \cite{BSSV}. Some years ago, the E704
Collaboration~\cite{E70495} at FNAL measured single spin
asymmetries for direct photon production in $pp$ collisions at 200
GeV/c. Although the single spin asymmetry for the direct-photon
production was found consistent with zero, within the experimental
uncertainty, there is nowadays a real possibility to increase the
precision of the measurement. In this letter, we show how to
relate the transverse SSA to the gluon Sivers function.

There are only two hard scattering processes for the direct photon
production in high $p_T$ collisions. One is the lowest-order
Compton subprocess, $qg \to \gamma q$ and the other one is the
lowest-order annihilation subprocess, $q \bar q \to \gamma g$.
However, since the first subprocess is dominant in $pp \to \gamma
X$ collisions, the unpolarized cross section for producing a
photon of transverse momentum $p_T$ and rapidity $y$ can be
written approximately as

\begin{eqnarray}
\nonumber
d\sigma&=& \sum \limits_{i} \int_{x_{min}}^1  dx_a
\int d^2{\bf k}_{Ta} d^2{\bf k}_{Tb}
\frac{x_a x_b}{x_a -({p_T /\sqrt{s})e^y }}
\left [ q_i(x_a,{\bf k}_{Ta}) G(x_b,{\bf k}_{Tb}) \right.\\
& &\times \frac{d\hat{\sigma}}{d \hat{t}}
 (q_i G \to q_i \gamma)
 + \left.  G(x_a,{\bf k}_{Ta}) ~q_i (x_b,{\bf k}_{Tb}) \frac{d\hat{\sigma}}{d
\hat{t}} ( G q_i \to q_i \gamma  ) \right ],
\end{eqnarray}
where $q_i(x,{\bf k}_{T})$ [ $G(x,{\bf k}_{T})$ ] is the quark [gluon]
distribution function with specified ${\bf k}_{T}$. A priori $k_T$, the
magnitude of ${\bf k}_{T}$, is
expected to be small compared to $\sqrt{s}$, where $s$ is the center of mass
energy of the reaction $pp \to \gamma X$. Therefore in order to simplify
our discussion, we will use the following expressions
\begin{equation}
x_b = \frac{x_a (p_T/\sqrt{s})~e^{-y}}{x_a -(p_T/\sqrt{s})~e^{y}}~,~~
x_{min}= \frac{(p_T/\sqrt{s})~e^{y}}{1 -(p_T/\sqrt{s})~e^{-y}}~,
\end{equation}
which are valid only in the collinear approximation.
The subprocess cross section is

\begin{equation}
\frac{d\hat{\sigma}}{d \hat{t}}
 (q_i G \to q_i \gamma  )=-\frac{\pi e_q^2\alpha \alpha_s }{3\hat{s}^2} \left [
\frac{\hat{u}}{\hat{s}}+
 \frac{\hat{s}}{\hat{u}} \right ],
\end{equation}
and by replacing $\hat{u}$ by $\hat{t}$, one obtains the other
internal cross section occurring in Eq.~(1). Here $\alpha$ is the
fine structure constant, $\alpha_s$ is the strong coupling
constant, $e_q$ denotes the quark charge and $\hat{s}$, $\hat{t}$,
$\hat{u}$ stand for the Mandelstam variables for the parton
subprocess

\begin{equation}
\hat{s}=x_a x_b s, \hspace{0.5cm} \hat{u}= - x_a p_T \sqrt{s}e^{-y},
\hspace{0.5cm}
\hat{t}= - x_b p_T \sqrt{s}e^{y}.
\end{equation}

According to the general definition of the ${\bf k}_T$-dependent
parton distributions $f(x,{\bf k}_T)$ ($f=q$, $G$) inside a
transversely polarized proton, where spin-{\it up} is labeled with
$\uparrow$ and {\it down} with $\downarrow$, it is clear that
\begin{eqnarray}
f(x,{\bf k}_T)&=&
\frac{1}{2}[f_{\uparrow}(x,{\bf k}_T) + f_{\downarrow}(x,{\bf k}_T)]
\nonumber \\
&=& \frac{1}{2}[f_{\uparrow}(x,{\bf k}_T) + f_{\uparrow}(x,-{\bf
k}_T)]=
 f(x,k_T) ,
\end{eqnarray}
whereas for the Sivers functions~\cite {DS} we have

\begin{eqnarray}
\Delta f_N(x,{\bf k}_T)&=&
f_{\uparrow}(x,{\bf k}_T) -  f_{\downarrow}(x,{\bf k}_T)
\nonumber \\
&=& f_{\uparrow}(x,{\bf k}_T) -  f_{\uparrow}(x,-{\bf k}_T)=
\Delta f_N(x,k_T){\bf S}_p\cdot{\bf \hat{p}}\times {\bf k}_T~~.
\end{eqnarray}
Here ${\bf S}_p$ denotes the transverse polarization of the
proton of three-momentum ${\bf p}$ and ${\bf \hat {p}}$ is a unit vector
in the direction of ${\bf p}$. The correlation proposed by
Sivers corresponds to a time-reversal odd triple vector product.
 Now we can define the SSA as

\begin{equation}
A_N^{\gamma} =\frac{d \Delta_
N \sigma }{d\sigma}~,
\end{equation}
where
$d\Delta_N \sigma = d\sigma^{\uparrow} - d\sigma^{\downarrow}$, whereas
$d \sigma = d\sigma^{\uparrow} + d\sigma^{\downarrow}$ and we have
\begin{eqnarray}
\nonumber
d\Delta_N \sigma &=&\sum \limits_{i}\int_{x_{min}}^1  dx_a \int
d^2{\bf k}_{Ta} d^2{\bf k}_{Tb}
\frac{x_a x_b}{x_a - (p_T/\sqrt{s})~e^y}
\left [ q_i(x_a,{\bf k}_{Ta}) \Delta_N G(x_b,{\bf k}_{Tb}) \right. \\
& & \times \frac{d\hat{\sigma}}{d \hat{t}}
 (q_i G \to q_i \gamma )
 +  \left.  G(x_a,{\bf k}_{Ta}) \Delta_N q_i (x_b,{\bf k}_{Tb})
 \frac{d\hat{\sigma}}{d \hat{t}} ( G q_i \to q_i \gamma  ) \right ]~~.
\end{eqnarray}
A priori the ${\bf k}_T$-dependence of all these parton
distributions is unknown, but as an approximation one can assume a
simple factorized form for the distribution functions and take for
example, as in Ref.~\cite{SS03},
\begin{equation}
f(x,k_T)= f(x)\lambda (k_T),
\end{equation}
where $\lambda (k_T)$ is flavor independent, and a similar
expression for the corresponding Sivers functions

\begin{equation}
\Delta_N f(x,k_T)= \Delta_N f(x)\eta (k_T)~.
\end{equation}

In such a situation \footnote{The simplifying assumptions
used above for the kinematics in the collinear approximation (see Eq.~(2)),
is justified by taking Gaussian expressions for $\lambda(k_T)$ and $\eta (f_T).$}, it
is clear that the SSA will also factorize and then it reads
\begin{equation}
A_N^{\gamma} (s, x_F, {\bf p}_T)= H(p_T)A^{\gamma}(s,x_F){\bf S}_p
\cdot{\bf \hat{p}}\times {\bf {p}}_T,
\end{equation}
where ${\bf p}_T$ is the transverse momentum of the photon
produced at the c.m. energy $\sqrt{s}$, and $H(p_T)$ is a function
of $p_T$, the magnitude of ${\bf p}_T$. We also recall the well
known relation between $y$ and $x_F$, namely
$x_F=2~\mbox{sinh}y(p_T/\sqrt{s})$.

\vspace{0.3cm}
\begin{figure}[htb]
\begin{center}
\leavevmode {\epsfysize=6.5cm \epsffile{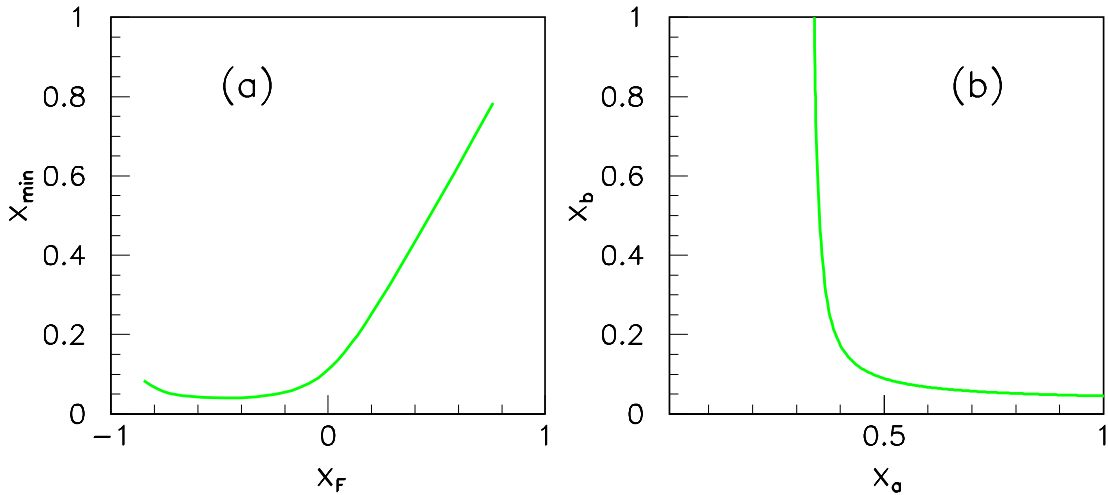}}
\end{center}
\caption[*]{\baselineskip 13pt For $\sqrt{s}=200~\rm{GeV}$, $p_T=20~\rm{GeV}$:
(a)
 $x_{min}$ versus $x_F$ and (b) $x_b$ versus $x_a$.
}\label{ssy1f1}
\end{figure}

Both Sivers functions for quarks and gluons are involved in
$A^{\gamma}(s,x_F)$, and therefore we want to identify a kinematic
region where the gluon Sivers function dominates. To achieve that
it is necessary to determine in Eqs.~(1) and (8), the range of
integration over $x_a$ and to study the relative magnitude of
$x_a$ and $x_b$. As an example, using Eq.~(2) with
$\sqrt{s}=200~\rm{GeV}$ and $p_T=20~\rm{GeV}$, the results for
$x_{min}$ versus $x_F$ are shown in Fig.~\ref{ssy1f1}(a) and we
find that $x_{min}\approx x_F$ in the region $x_F > 0.3 $. On the
other hand, $x_b$ versus $x_a$ is shown in Fig.~\ref{ssy1f1}(b) and
we see that when $x_a$ is integrated over the range [$x_{min}$,
1], the main contribution comes from the low $x_b$ values.
Therefore, when we look at the large $x_F$ region, where $x_a$ is
large but $x_b$ is small, the asymmetry can be approximately
expressed as

\begin{equation}
A^{\gamma}(s,x_F) = \frac{ \langle \Delta_N G \rangle }
{ \langle G \rangle }~,
\end{equation}
where $\langle \Delta_N G \rangle$ and $\langle G \rangle$ mean
the corresponding values over an appropriate integrating range.
Unlike the quark Sivers functions, for which several theoretical
calculations have been performed, for example in a spectator model
with axial-vector diquarks (see Ref.~\cite{BSY03} and references
therein), the gluon Sivers function have not been really
investigated, so we will not try to use a numerical estimate for
$\Delta_{N}G$. On the experimental side the inaccurate result of
Ref.~\cite{E70495} is anyway irrelevant for our purpose, because
it concerns the central region $x_F \sim 0$. On the other hand it
is worth mentioning the measurement of the SSA in the very forward
production of photons in $pp$ collisions at $\sqrt{s}=200
\mbox{GeV}$ with $p_T<<0.5 \mbox{GeV}$, consistent with zero \cite{BAZ}.
The fact that they measure {\it all} photons and not only direct
photons, makes these data also irrelevant. This forward kinematic region 
is indeed quite accessible at
RHIC, since the PHENIX Collaboration has already released the
unpolarized cross section for $pp \to \gamma X$ at $\sqrt{s}=200
\mbox{GeV}$, in the central region for $p_T$ up to 18 GeV
\cite{KR}, in fair agreement with NLO pQCD calculations. The same
calculation predicts for $p_T \sim 8 \mbox{GeV}$ and $x_F \sim
0.3$, a cross section of about $40\mbox{pb}/\mbox{GeV}^2$
\cite{WV}. We hope this will be a good motivation to undertake
the measurement of the SSA, but we know that the extraction of the
gluon Sivers function, even if it turns out to be large, will not
be straightforward. Among the various effects which might dilute
the SSA, it is important to mention the effects of QCD gluon
resummation \cite{NSY,DB} and Sudakov effects have been shown to
lead to significant suppression of the SSA considered in Ref.~\cite{BV}.

Other similar processes are $p p^{\uparrow} \to \gamma + jet + X$,
muon pair production $p p^{\uparrow} \to \gamma^* + X \to \mu^+
\mu^- + X$ and $\bar{p} p^{\uparrow} \to \gamma + X$. The first
reaction is certainly very interesting also, because by detecting
simultaneously the photon and the jet, one has both rapidities to
consider and Eq.~(12) becomes simpler, with no integrations.
For muon pair production, the outgoing photon is monitored by its
conversion to muon pairs and this process is more difficult to
study experimentally. Finally, in the case of $\bar{p}
p^{\uparrow} \to \gamma + X$, the quark annihilation process
$\bar{q} q \to \gamma g$ dominates, which makes it unpractical.
Therefore, the ideal probe to extract the gluon Sivers function is
the transverse single spin asymmetry of prompt photon production
at high $p_T$, and RHIC is obviously very suitable to realize this
important measurement with good precision.

{\bf Acknowledgments: } This work is partially supported by
Fondecyt (Chile) under Grant Number 1039355 and by the cooperation
programme Ecos-Conicyt C04E04 between France and Chile. We thank G. Bunce
for useful comments.


\begin{thebibliography}{99}

\bibitem{E704}E704 Collaboration, D. L. Adams {\it et al.}, Phys. Lett. {\bf B
261} (1991) 201
and Phys. Lett. {\bf B 264} (1991);
A. Bravar {\it et al.}, Phys. Rev. Lett. {\bf 77} (1996) 2626.

\bibitem{STAR} STAR Collaboration, J. Adams  {\it et al.}, Phys. Rev. Lett. {\bf
92} (2004)
171801.

\bibitem{BS} C. Bourrely, J. Soffer, Eur. Phys. J. {\bf C 36} (2004) 371.

\bibitem{LP} L. Pondrom, Physics Reports {\bf 122} (1985) 57; K. Heller,
 Proceedings of "Spin96", Amsterdam 10-14/10/96, World Scientific
(1997) p.23; for recent
 results see also Proceedings of "Hyperon99", FNAL Batavia
27-29/09/99, FERMILAB-Conf-00/059-E
 (Eds. D. A. Jensen and E. Monnier).

\bibitem{SMC} SMC Collaboration, S. Bravar,  Nucl. Phys. B (Proc. Suppl.) {\bf
79} (1999) 520.

\bibitem{hermes} HERMES Collaboration, A. Airapetian {\it et al.}, Phys. Rev.
Lett.
{\bf 84} (2000) 4047; Phys. Rev. {\bf D 64} (2001) 097101;
 Phys. Lett. {\bf B 562} (2003) 182.

\bibitem{BHS1}S. J. Brodsky, D. S. Hwang and I. Schmidt,
 Phys. Lett. {\bf B 530} (2002) 99.

\bibitem{QS92} J. Qiu and G. Sterman, Phys. Rev. Lett. {\bf 67} (1991) 2264;
Nucl. Phys. {\bf B 378} (1992) 52.


\bibitem{DS} D. Sivers, Phys. Rev. {\bf D 41} (1990) 83 and
 Phys. Rev. {\bf D 43} (1991) 261.

\bibitem{JC}J. C. Collins, Nucl. Phys. {\bf B 396} (1993) 43.

\bibitem{BSY03} A. Bacchetta, A. Sch\"afer, J. J. Yang,
 Phys. Lett. {\bf B 578} (2004) 109.

\bibitem{SS03} I. Schmidt and J. Soffer, Phys. Lett. {\bf B 563} (2003) 179.

\bibitem{BV} D. Boer, W. Vogelsang, Phys. Rev. {\bf D 69} (2004) 094025.

\bibitem{ABALM} M. Anselmino {\it et al.}, Phys. Rev. {\bf D 70} (2004) 074025.

\bibitem{BSSV} G. Bunce, N. Saito, J. Soffer, W. Vogelsang, Annu. Rev. Nucl.
Part. Sci. {\bf 50} (2000) 525.

\bibitem{E70495} E704 Collaboration, D. L. Adams {\it et al.}, Phys. Lett. {\bf B 345} (1995) 569.

\bibitem{BAZ} A. Bazilevsky {\it et al.}, presented at "`SPIN 2002"', AIP Conference Proceedings 675, p.584 (2003).

\bibitem{KR} PHENIX Collaboration, K. Okada {\it et al.},  hep-ex/0501066, contribution to SPIN2004,
Trieste, Italy, Oct. 10-16 (2004) to appear in the proceedings.

\bibitem{WV} We thank W. Vogelsang for providing us with this numerical value.

\bibitem{NSY} P. Nadolsky, D. Stump and C.P. Yuan, Phys. Rev. {\bf D 61} (2000) 014003
and Phys. Rev. {\bf D 64} (2001) 114011.

\bibitem{DB} D. Boer, Nucl. Phys. {\bf B 603} (2001) 195.
\end{thebibliography}
\end{document}